\documentstyle[12pt,aaspp4]{article}
\begin{document}

\title{MASS DETERMINATION WITH GRAVITATIONAL MICROLENSING}
\author{Shude Mao$^1$ and Bohdan Paczy\'nski$^{2}$}
\affil{$^1$ Max-Planck-Institute f\"ur Astrophysik,
Karl-Schwarzschild-Strasse 1, 85740 Garching, Germany}
\affil{E-mail: smao@mpa-garching.mpg.de}
\affil{$^2$Princeton University Observatory, Princeton, NJ 08544--1001, USA}
\affil{E-mail: bp@astro.princeton.edu}


\begin{abstract}

We present a simple toy model of the distribution of objects responsible
for gravitational microlensing.  We use Monte Carlo simulations
to demonstrate how difficult it is to determine the parameters
of the lens mass distribution on the basis of the observed distribution
of event time scales.  A robust determination requires $ \sim 100 $
events, or more, even if the geometry of lens distribution,
and the lens kinematics are known.

\end{abstract}

\section{Introduction}

One of the main objectives of the searches for microlensing
events (Paczy\'nski 1996, and references therein)
is to determine the mass function of the lensing objects.
The determination of typical lens masses, or even their
distribution function, were attempted by
Alcock et al. (1993), Udalski et al. (1994), Zhao
et al. (1995, 1996), Han \& Gould (1996), and many others.

We know that some lenses must be ordinary stars, but some
may be brown dwarfs, and perhaps also planetary mass objects,
and even more exotic things like black holes.  It is recognized
that even if all lensing objects were of the same mass $ M $,
they would give rise to a broad range of time scales $ t_0 $
of the observed microlensing events.
Although this is known to a relatively small number
of people in the field this fact is not generally appreciated.
Considering the broad interest in the nature of dark matter,
and the relevance of microlensing searches to this problem, 
we feel it is justified to present a very simple model
of the lens distribution and kinematics
which allows some results to be obtained analytically, and
the rest can be calculated with one-dimensional numerical
integrals.  This model retains some of the most important generic
characteristics: a very broad distribution of $ t_0 $ 
with power law tails towards very short and very long time scales
for a delta function distribution of lens masses.
Using Monte Carlo simulations we demonstrate how
difficult it is to obtain accurate information about the
mass function even within the framework of our simple model.
In reality one is not sure what is the correct
space distribution and kinematics of lensing masses,
making the interpretation even more difficult.

The model is described in the next section.  The results of
the Monte Carlo simulations are presented in section 3, and
the discussion of the results is given in the last section.
The technical details of the model can be found in the appendix.

\section{A Model}

We consider a simple case that may approximate a real astrophysical
system: the lenses, each of the same mass $ M $, have their distances
$ D_d $ distributed uniformly and randomly
between the observer and a stationary
source, which is at a distance $ D_s $.
The time scale of a microlensing event is given by
$$
t_0 = { R_E \over V_t } ,
\hskip 1.0cm V_t \equiv V \sin i ,           \eqno(1a)
$$
where $V$ is the relative spatial velocity between the lens and the
observer, $ i $ is the angle between the velocity vector and the line of
sight, and $ R_E $ is the Einstein ring radius:
$$
R_E = \left( 2 R_g D \right) ^{1/2} ,  \hskip 1.0cm
R_g = { 2GM \over c^2 } , \hskip 1.0cm
D = { (D_s - D_d) D_d \over D_s } .                      \eqno(1b)
$$
More technical details about our model are given in the appendix.
In this section we present only the most essential information.

The lenses in our model have a three-dimensional Gaussian velocity 
distribution with a one-dimensional dispersion of
$ \sigma _V $.  A characteristic
time scale for a microlensing event is defined as
$$
t_{ \sigma} = { 1 \over \sigma _V } ~
\left( { GMD_s \over c^2 } \right) ^{1/2}
 = 70 ~ {\mathrm{days}} ~ 
\left( { 100 ~ {\mathrm{km ~ s^{-1}}} \over \sigma _V } \right) ~
\left( { M \over 1 ~ M_{\odot} } \right) ^{1/2} ~
\left( { D_s \over 8 ~ {\mathrm{kpc}} } \right) ^{1/2} ~ . \eqno(2)
$$

The probability distribution for event time scales has power law tails:
$$
P_\delta(<t) = { 2^{11/2} \over 3 \pi^{3/2} } ~ t^3, \hskip 1.0cm
{\mathrm{for}} ~~ t \ll 1 ,  \eqno(3a)
$$
$$
P_\delta(>t) = { 2^{11/2} \over 45 \pi^{3/2} } ~ {1 \over t^3}, \hskip 1.0cm
{\mathrm{for}} ~~ t \gg 1 ,  \eqno(3b)
$$
where
$$
t \equiv t_0/t_{\sigma},    \eqno(3c)
$$
and the subscript $\delta$ indicates that all the lenses have the
same, i.e., we have a $\delta$ mass function.
The full probability distribution of event time scales 
can be calculated numerically and it can be
approximated within 6\% with a formula
$$
p_\delta (\log t) ~ d \log t \approx  { c t^3 \over 1 + 15 t^6 } \times
10^{ -1.44 ( \log t +0.293) \exp [ - (( \log t + 0.28 )/0.304)^2 ] } 
~ d \log t , 
\eqno(4)
$$
where
$$
c = { 2^{5.5} \over \pi ^{1.5} } \ln 10 \approx 18.71 ,
$$
The first two moments and the standard deviation of this distribution are 
$$
<\log t>_\delta \equiv \int_0^{\infty} p_\delta(t)~\log t~dt = -0.242 , \eqno(5a)
$$
$$
< ( \log t )^2 >_\delta \equiv \int_0^{\infty} p_\delta(t)~ ( \log t )^2~dt = 0.116 .
\eqno(5b)
$$
$$
\sigma _\delta \equiv 
\left[ < ( \log t )^2 >_\delta - \left( <\log t>_\delta \right) ^2 \right] ^{1/2} 
= 0.240 .  \eqno(5c)
$$

Naturally, formula (4) applies only to our simple model of the lens
distribution and kinematics.  However, the presence of power law tails
is common to almost all distributions ever proposed (e.g., those
in Zhao et al. 1996) and
can be understood intuitively. The very long
events are produced by the lenses which move almost along the line of sight.
For these events the time scale $t$ is inversely proportional 
to $\sin i$, i.e., $ t \propto i^{-1} $ (cf. eq. 1a).
The number of these events, $N$, is proportional
to the solid angle $(1-\cos i)$ divided by $t$, i.e.,
$ N \propto (1-\cos i)/t \propto i^3 \propto t^{-3}$. 
On the other hand, the short events are
produced by the lenses which are very close to either the source or
the observer.  For the latter case,
$t \propto R_E \propto D_d^{1/2}$, and the 
number of these events is $ N \propto R_E^2 D_d/t \propto t^3$. 
The same argument applies when the lenses are close to the source.

It is not likely that all the lenses have identical masses, and it is
convenient to consider a power law mass distribution
$$
p ~ (M) ~ d \left( {M \over M_0} \right) = 
A \left( {M \over M_0 } \right) ^{ \alpha } ~ 
d \left( {M \over M_0 } \right) ,
\hskip 1.0cm {\mathrm{for}} \hskip 0.5cm M_{\mathrm{min}} \leq M \leq M_{\mathrm{max}},
\eqno(6a)
$$
where
$$
M_0 = ( M_{\mathrm{min}} M_{\mathrm{max}} )^{1/2},
\eqno(6b)
$$
and we denote the logarithmic width of the mass function by
$$
\beta \equiv \log ~ (M_{\mathrm{max}}/M_{\mathrm{min}}). \eqno(6c)
$$
The exponent $ \alpha = -1.5 $ corresponds to an
equal rate of microlensing events per decade of lens masses, while
the case $ \alpha = -2 $ corresponds to an equal contribution
to the optical depth per decade of lens masses. For
$-2<\alpha<-1.5$, the optical depth (total mass) is dominated by the massive
objects, while the event rate is dominated by the low mass objects.
For a given mass range the standard deviation of $ \log t $ is the
largest for $ \alpha = -1.5 $. 
For $ \alpha \ll -1.5 $ the mass spectrum
is dominated by low mass objects, and the event time scales
have a small effective range, with the majority of events caused
by lenses with $ M \approx M_{\mathrm{min}} $.  In the opposite case, with
$ \alpha \gg -1.5 $, the microlensing events are dominated by lenses
with $ M \approx M_{\mathrm{max}} $.
Therefore for $\alpha > -1.5$, the standard deviation in $\log t$
approaches some asymptotic value when $M_{\mathrm{min}} \rightarrow 0$, and
similarly for $\alpha < -1.5$ when $M_{\mathrm{max}} \rightarrow \infty $.
After some algebra one may show that the variance in the observed
distribution of event time scales can be expressed as
$$
\sigma _{ \log t }^2 = 
\sigma_\delta^2 + { \beta ^2 \over 48 }, 
\hskip 1.0cm {\mathrm{for}} \hskip 0.5cm \alpha = -1.5 ,
\eqno(7)
$$
and the asymptotic value is
$$
\sigma^2_{\log t, \infty} = \sigma^2_\delta + 
\left[ 2 ~(\alpha +1.5) ~ \ln 10 \right] ^{-2} ,
\hskip 1.0cm {\mathrm{for}} \hskip 0.5cm \alpha \ne -1.5 ,
\hskip 0.5cm \beta \rightarrow \infty ,  \eqno(8)
$$
where $ \sigma _\delta = 0.240 $, as given by eq. (5c).

The probability distribution functions for the logarithm of event time
scale for any lens mass function
can be calculated by convolving the mass function with the probability
distribution for a $\delta$ mass function as given in eq. (4).
The probability distributions for
$ ( \alpha , \beta ) = (-1.5,1), ~ (-1.5,2), ~ (-2.5,1), ~ (-2.5,2) $,
are shown in Fig. 1. All time scales are in units of $t_\sigma $
corresponding to $ M = M_0 \equiv (M_{\mathrm{min}} M_{\mathrm{max}})^{1/2} $ 
(cf. eqs. 2 and 6).
Also shown is the case for $ \beta = 0 $, i.e., with all the lenses having
identical masses (cf. eq. 4). The shapes for mass functions
with a narrow width ($\beta \la 1$) are quite similar.
This can be easily understood because the time scale of microlensing
$t_0 \propto M^{1/2}$, therefore one decade of mass range translates
to a factor of three in the time scale, hence the shapes
of these distributions still resemble that for a $\delta$ mass function.

\section{Monte Carlo Simulations}

We present two types of Monte Carlo simulations.
In both cases we adopt
the model of lens distribution and kinematics as developed in the previous
section.
First, we consider a
perfect detection system, which can detect every event of any time scale,
and a lens population with a moderately broad power law mass function.
Next we consider very broad lens mass functions and a detection system
which is like that of the MACHO and OGLE collaborations, i.e., it has
a significant detection efficiency over a relatively narrow range of
event time scales.  In both cases we make a large number of the
Monte Carlo simulations to determine the accuracy with which the lens mass
function parameters can be estimated with a moderately large number
of detected events: $ n = 10 $, $ n = 100 $, and $ n = 1,000 $.

We tried both the moment method and the maximum
likelihood method to estimate the parameters $ \alpha $ and $ \beta $.
Although the moment analysis is faster and in many cases its results agree
with that from the maximum likelihood, it is less accurate
and in some cases does not work at all. One of the reasons is
very easy to understand.
Imagine we have a random sample of microlensing events with
variance smaller than $ \sigma _{ \delta } = 0.240 $, as given
by eq. (5c).  Such cases are unavoidable in a finite sample
drawn from a distribution with $ \sigma > \sigma _{ \delta } $. 
The moment analysis fails for these cases.
This reminds us of a similar old problem: what to
do with negative trigonometric parallaxes of distant stars, the
negative numbers being due to the unavoidable measurement errors?
For this reason, we only present the results from the
maximum likelihood method with a prior of equal probability
in $\alpha$ and $\beta$. However, general discussions are given using
the more intuitive moment analysis.

\subsection{Finite Mass Range, Perfect Detection System}

We adopt a power law lens mass function described with two dimensionless
parameters: the power law exponent $ \alpha $, and the range of 
masses $ \beta \equiv \log ~ (M_{\mathrm{max}}/M_{\mathrm{min}}) $ (cf. eq. 6).
In addition, we have two scaling parameters: $ A $ and $ M_0 \equiv
( M_{\mathrm{min}} M_{\mathrm{max}} )^{1/2} $ (cf. eq. 6). Other things being equal the
optical depth and the rate of microlensing events are proportional
to $ A $, while the average time scale of the events is proportional
to $ M_0^{1/2} $. Out of the four parameters,
the scaling parameter $A$ can be fixed by the observed event rate,
after which we are still left with three parameters. We can simplify
the problem further by identifying a mass scale in the problem. In 
most cases, we assume such a scale is provided by a fixed upper mass
limit, say $M_{\mathrm{max}} = 1 M_\odot$, but we also explore another possibility
(see below).

We assume that the lens detection system is perfect, i.e., it is 100\%
efficient for all event time scales.  A large number of Monte Carlo
samples are generated
with the {\it average} number of events $ n = 100 $ and $ n = 1,000 $.
The actual number of events was chosen from a Poisson distribution,
followed by a Monte Carlo selection of event time scales according
to the model probability distributions, as shown in Fig. 1.

We selected two models with
$ ( \alpha , \beta ) = (-1.5,1), ~ (-1.5,2)$, each for two sample sizes:
$n=100$ and $n=1000$. The $1\sigma$ and $2\sigma$ contours are shown
in Fig. 2. It is clear that for the case of $(\alpha, \beta)=(-1.5, 1)$, the
power law index $\alpha$ cannot be determined very well
with 100 events, the case for $(\alpha, \beta)=(-1.5,2)$, however, is
improved. The large difference between $\beta=1$ and $\beta=2$ is because
for $\beta=1$, the shapes, and therefore the variances, of the duration
distributions for different power law indices are similar (cf. Fig. 1).
The main difference between the models is the average duration, i.e.,
the first moment of the duration distribution. However,
such difference can be accommodated by slightly changing the lower mass
limit. The power law index $\alpha$ is then mostly constrained by the
more uncertain higher moments, and hence the weak limit.

The case with a broader mass function, $ \beta = 2 $, is better constrained
than that for $\beta=1$ with a sample of $ n = 100 $
microlensing events, but even this case
is very poorly constrained when $ n = 10 $, as shown with the dotted
contours in Fig. 2.  Note that the total number of microlensing
events detected so far towards the LMC is smaller than 10 (Bennett et al.
1996) and the
spatial distribution and kinematics of the lenses are not known, making
the case even more ill-constrained.

The contours shown in Fig. 2 are based the assumption that
the upper mass limit is known. Such a fixed upper limit 
sets the mass scale in the maximum likelihood method and
translates the observed average duration (i.e., the first moment of the
duration distribution) into a constraint on the lower mass limit.
In reality, we do not know the upper mass limit well. In fact, this
upper limit should also be found from the maximum likelihood. However,
it is too time consuming to conduct such a three dimensional study.
Instead we adopt a different approach: for a given set of $(\alpha, \beta)$,
we fix the mass scale such that it gives identical
first moment in the duration distribution as a generated Monte Carlo sample.
This reduces the three-dimensional problem into a two dimensional
one, and should be a reasonable approximation. As the
first moment has been utilized to set the mass scale, the determination
of $ \alpha $ and $ \beta $ is then effectively based on the second and
higher moments, and these are subject to larger random uncertainties.
The likelihood contours shown in Fig. 2 confirms
this for $\alpha=-1.5, \beta=2$ with $n=100$ (dashed
lines). Just as expected, the limit is 
relaxed -- the allowed parameter space is increased by a factor of
$ \approx 2 $.
This is a consequence of the determination made with the maximum likelihood
(or any other) method effectively using the higher moments of the event
distribution.

With the Monte Carlo samples, we can also estimate the optical depth of
microlensing, $ \tau $, which is proportional to the number
of events observed per unit time per unit number of stars.  The error
of the optical depth estimate is inversely proportional to the square 
root of the number of events, $ n^{-1/2} $, with a coefficient somewhat
larger than unity, as the relatively few longest events are most important.
This coefficient depends on the values of $ \alpha $ and $ \beta $.
For example, the coefficients are 1.3, 1.5, 1.3, 1.6
for $(\alpha, \beta)=(-1.5, 1), (-1.5, 2), (-2, 1), (-2, 2)$, respectively.
This is in good agreement with Han \& Gould (1995).
If we have 100 microlensing events, and if we can trust our model of the 
lens distribution and kinematics, then we can estimate the value of 
$ \tau $ with an accuracy of $ \approx 15\% $.

\subsection{Very Broad Mass Range, Imperfect Detection System}

Now we consider a very broad mass function with $ \beta \gg 1 $,
so the time scales of microlensing events cover
the range $ \Delta \log t_0 \approx 0.5 \beta $.  We
adopt the efficiency of event detectability similar to those published
by the OGLE (Udalski et al. 1994) and MACHO (Alcock et al. 1995a)
collaborations.  We approximate the detection efficiency
with a  formula:
$$
\epsilon = 
\left\{		\begin{array}{ll}
		0.4 \exp[-(\log t)^2/0.72],  &
		\mbox{if $\log t < 0$} \\
		0.4 \exp[-(\log t)^2/0.18],  &
		\mbox{if $\log t > 0$}
		\end{array}
\right.
\eqno(10)
$$
where $ t \equiv t_0/t_{p} $, and $ t_{p} \approx 35 $ days
corresponds to events with durations that are at the peak of
detection sensitivity. The detectability window as described
by eq. (10) has a width of about one decade at half maximum.
Using eq. (2) we may relate this time scale to a mass scale
$$
M_p = 0.25 M_{\odot}  ~
\left( { \sigma _V \over 100 ~ {\mathrm{km ~ s^{-1}}}} \right) ^2 ~
\left( { 8 ~ {\mathrm{kpc}} \over D_s } \right) .  \eqno(11)
$$

Again we adopt a power law mass function as given by eq. (6). It
is convenient to express the lower and upper mass limits in units
of $M_p$ with $\beta _{\mathrm{min}} \equiv \log ~ (M_{\mathrm{min}}/M_p) $
and $\beta _{\mathrm{max}} \equiv \log ~ (M_{\mathrm{max}}/M_p)$.
The probability distributions of event time scales for very broad mass
functions, with $ \beta = 10 $ and slopes $ \alpha = -2, ~ -1.5, ~ -1 $,
are shown in Fig. 3.  Also shown is 
the efficiency of event detection as described with eq. (10).

If the probability function of event time scales is much broader than
the width of the detectability window as given by eq. (10) then
most events cannot be detected.  We shall consider two cases.  First,
a mass function extending very far towards low masses, i.e., with
$ \beta _{\mathrm{min}} \ll -1 $, and the upper mass 
limit somewhat larger than $ M_p $.  Second, a mass function extending 
very far towards large masses, i.e., with $ \beta _{\mathrm{max}} \gg 1 $,
and the lower mass limit somewhat smaller than $ M_p $.  We shall 
investigate the ability of the detection system to measure 
$ \beta _{\mathrm{max}} $ in the first case, and 
$ \beta _{\mathrm{min}} $ in the second case.

In all the simulations we adopt the average number of detected events
to be either $ n = 100 $ or $ n = 1,000 $, the actual number being
drawn from a Poisson distribution.  
Assuming that $ \beta _{\mathrm{min}} \ll -1 $
we can search for the remaining two dimensionless parameters:
$ \alpha $ and $ \beta _{\mathrm{max}} $ using the maximum likelihood method,
with $A$ being fixed by the event rate.
The results of model calculations are shown in the right panel of
Fig. 4 for $ (\alpha, \beta_{\mathrm{max}}) = (-1.5, 1)$, and $(-1.5, 2)$.
With 100 events, $\alpha$
is well constrained in both cases. The parameter
$\beta_{\mathrm{max}}$, however, becomes not so well-determined
for $\beta_{\mathrm{max}}=2$.

Assuming that $ \beta _{\mathrm{max}} \gg 1 $
we can search for the remaining two dimensionless parameters:
$ \alpha $ and $ \beta _{\mathrm{min}} $ using the maximum likelihood method.
The results of model calculations are shown in the left panel of
Fig. 4 for ($ \alpha, \beta _{\mathrm{min}}) = (-1.5, -3), (-1.5, -2)$,
and $(-1.5, -1)$. The parameter $\alpha$ is again
well constrained for $n=100$. In comparison, the parameter
$ \beta_{\mathrm{min}}$ becomes ill-determined when 
it approaches $-3$.

Note that the models with $ \beta _{\mathrm{min}} = -3 $ and
$ \beta _{\mathrm{max}} = 2 $
have contours slightly shifted with respect to the true model values,
as marked with crosses.
If the distribution of event time scales is approaching the effective
edge of the detectability window then some finite samples of events
generate second (and other) moments which are incompatible with
the model. For such cases a method using the moments simply
fails. The maximum likelihood method can better handle such cases,
but has some trouble as well for $n=100$, as indicated by the small
shifts of contours in Fig. 4. This can be remedied with
a larger sample size such as $n=1000$ (thick dashed lines) in both cases.

Given the determination of $ \alpha $ and either $ \beta _{\mathrm{min}} $ or
$ \beta _{\mathrm{max}} $, as well as the `observed' number of events `n'
we may estimate the parameter $ A $ in eq. (6).  The error
of the estimate can be written as
$$
\sigma _A \approx a ~ n^{-1/2} , \eqno(12)
$$
where $ \sigma _A $ is the standard deviation in the
$ A_{\mathrm{sample}} / A_{\mathrm{model}} $ ratio.
The coefficient $ a $ depends somewhat on $ \alpha $ and $ \beta $,
but is generally in the range between 1 and 1.6.

\section{Discussion}

The Monte Carlo simulations presented in this paper demonstrate that
even if we have a full knowledge of the space distribution and
kinematics of the lensing objects, or equivalently, if we know the
relation between the lens mass and the distribution of event time scales
(see eq. 4 for an example), it may still be difficult to obtain the 
parameters of the lens mass function. The problem is particularly
serious when the width of the mass function is narrow ($\beta \la 1$).
In addition, if the mass function power law is quite different from 
$-1.5$, then it will be difficult to probe the high mass end 
when $\alpha \ll -1.5$, and the low mass end when $\alpha \gg -1.5$
(cf. eq. 8).

Recently, Han \& Gould (1996) obtained a rather tight limit on the mass
function of the Galactic disk using about 50 microlensing events toward
the Galactic bulge. In their analysis,
the upper mass limit is fixed at $10 M_{\odot} $ in a rather
ad hoc manner. As we have shown in Fig. 2, such a fixed upper mass
limit will make the determination of the other parameters appear
more accurate.
A more realistic approach is to estimate the upper mass
limit using the maximum likelihood in order to avoid carrying
our a priori assumption into a limit on the physical parameters.
In general, if at least three separate parameters are to be estimated,
for example $ M_0$, $ \alpha $, and $ \beta $, then any
method must effectively use the information contained in at least the
first three moments.  However, the higher the moment the more uncertain 
is its value as estimated from a relatively small sample of events.
Therefore, the accuracy of the determination of any distribution parameter
is lower in the 3-parameter determination than in a 2-parameter determination.
Unfortunately, the smaller errors in a 2-parameter determination could be
misleading, unless we have an independent and reliable information
about the value of the third parameter.

In our second example, very broad mass functions are sampled through a
realistic detectability window (cf. Fig. 4). The current published
sensitivity windows allow one to probe about three to four decades
of mass range and we find that
the ``amplitude'' $ A $ and the local slope of the lens mass function
$ \alpha $ can be determined reasonably well with $ \sim 100 $ events.
However, the estimate of the range of masses within the detectability
window, $ \beta _{\mathrm{min}} $ or $ \beta _{\mathrm{max}} $, may be very uncertain
with as many as $ n = 100 $ events, as shown in Fig. 4.

If the mass function is broader than the detectability window then
the total event rate or the total optical depth (or total mass)
cannot be measured as the majority of
events are outside the detectability window (Fig. 3), having either too
short or too long $ t_0 $.  This points
to the necessity of broadening the detectability window, a task easy
to accomplish in the near future. Some first attempts have already been
made by the EROS (Ansari et al. 1995) and MACHO collaborations (Bennett 
et al. 1996).

How can we be sure that we have reached the low and the high mass
end of the lens distribution?  In principle this is simple: we have
to broaden the detectability window and we should detect so many
events that the generic power law tails in the distribution (cf. eq. 3)
become apparent.  This may call for well over 100 events. Such a high
number of events is within reach for the Galactic Bulge, where the
rate appears to be very high (Udalski et al. 1994, Alcock et al. 1995b,c).
The determinations of any parameters of the distribution of lens masses
based on as few as 10 events are subject to very large uncertainties,
as shown with the dotted contours in Fig. 2.

In reality we do not know what the geometry of the lens distribution is,
and what is the lens kinematics.  While looking towards the Galactic
Bulge the majority of lenses may be in the Bulge itself (Kiraga \& Paczy\'nski
1994, Udalski et al. 1994, Paczy\'nski et al. 1994, Zhao et al. 1995, 1996),
or they may be in the disk (Alcock et al. 1995b,c).  While looking towards
the LMC the majority of lenses may be in our galaxy (Alcock et al. 1995a)
or in the LMC (Sahu 1994; Wu 1994).  Depending on the location and kinematics
the relation between the lens masses and the event time scale $ t_0 $
may be very different, and this leads to an additional major uncertainty
on top of purely statistical uncertainty discussed in this paper.
It seems rather difficult to disentangle all parameters one needs to describe
the distribution and kinematics of the lenses {\it and} their mass
function on the basis of the observed distribution of event time scale.

Fortunately, the future studies of the variation of the optical depth
with the location in the sky will help to identify the lens location.
If lensing of the Galactic Bulge stars is dominated by the Bulge lenses
then the optical depth should vary rapidly with the galactic longitude
(Kiraga \& Paczy\'nski 1994).  If such a variation is not present this 
will indicate that the lenses are located mostly in the galactic disk.
If lensing towards the LMC is dominated
by the LMC lenses then the optical depth should increase strongly
towards the LMC center (Sahu 1994).  If the optical depth is uniform
over the LMC then the lenses must be mostly in our galaxy.  This is very
simple in principle, but it will take hundreds of events to establish
beyond reasonable doubt.  Such a large number will be readily collected
towards the Galactic Bulge, but it will take a very long time for
the LMC as the rate in that direction is very low (Alcock et
al. 1995a).  Once the geometry of the lens distribution is established
it will be possible to develop reliable models of their kinematics,
and infer a reasonable statistical relation between the lens masses
and the event time scales $ t_0 $.

\vskip 0.5cm

This project was supported by the NSF grant AST93-13620 and 
by the ``Sonderforschungsbereich 375-95 f\"ur Astro-Teilchenphysik''
der Deutschen Forschungsgemeinschaft. We are very grateful to
Dr. Peter Schneider for a critical reading of the manuscript.

\newpage
\appendix
\section{Appendix}

We begin with the simplest lens model: all lensing objects have the
same mass $ M $, the same three-dimensional velocity $ V $, and
their velocity vector directions have a random isotropic distribution.
The source located at the distance $ D_s $ is stationary, and the
number density of lensing objects is uniform between the observer
and the source. 

If all the events had identical time scales, then the number of microlensing
events expected in a time interval $ \Delta t $ would be given as
$$
N = { 2 \over \pi } ~ n_s \tau ~ { \Delta t \over t_0 } ,  
\hskip 1.0cm \mathrm {for} \hskip 0.5cm t_0 = {\mathrm{const}} ,  \eqno(A1)
$$
where $ 2 / \pi $ is the ratio of Einstein ring diameter to its area, in
dimensionless units, $n_s$ is the number of sources monitored,
and $ \tau $ is the optical depth of microlensing.
In fact there is a broad distribution of event time scales as lenses have
transverse velocities in the range of $ 0 \leq V_t \leq V $, and
distances in the range $ 0 \leq D_d \leq D_s $.  A straightforward but
tedious algebra leads to the equation
$$
N = { 3 \pi \over 16 } ~ n_s \tau ~ { \Delta t \over t_{m} } = N~
\int _0^{\infty} p_0(t_0) ~ dt_0 ,                     \eqno(A2a)
$$
where
$$
t_{m} \equiv
\left( { R_E \over V } \right) _{D_d=0.5 D_s} =
\left( { GMD_s \over c^2 } \right) ^{1/2} { 1 \over V } ,   \eqno(A2b)
$$
is the time scale for a microlensing event due to a lens located
half way between the source and the observer, and moving with the
transverse velocity $ V $, and the probability distribution
of event time scales is given by 
$$
p_0(t_0)~dt_0 = p_T(T) dT \equiv {1 \over \pi^2 T^3}
\left[ -6 T(1+T^2) + (3+2 T^2 + 3 T^4) 
\ln \left|{1+T \over 1-T} \right| \right]~ dT,   \eqno(A3a)
$$
where
$$
T \equiv {t_0 \over t_m}, \hskip 1.0cm 0<t_0<\infty.    \eqno(A3b)
$$
The probability distribution diverges
logarithmically at $t_0=t_m$,
however, the integrated probability converges. Also note
that the probability density distribution satisfies: $p_0(1/T) = T^2 p_0(T)$.
Therefore, the probability that the event time scale is longer than $ T(>1)$
is equal to the probability that the time scale is shorter than $T^{-1}$.

The asymptotic behaviors of the integrated probability are given by
$$
P_0(<T) = {128 \over 45 \pi^2} T^3, \hskip 1.0cm
{\mathrm{for}} ~~ T \ll 1 ,  \eqno(A4a)
$$
$$
P_0(>T) = {128 \over 45 \pi^2} {1 \over T^3}, \hskip 1.0cm
{\mathrm{for}} ~~ T \gg 1 . \eqno(A4b)
$$
The power law tails are generic to almost all lens distributions
ever proposed. Also note that
only the first two moments of the distribution are finite:
$ \langle t_0 \rangle $ and $ \langle t_0^2 \rangle $, while
the third and higher moments diverge.
Therefore, it is convenient to use a logarithmic probability 
distribution defined as
$$
p_0( \log t_0 ) ~ d \log t_0 =
( \ln 10 ) ~ t_0 ~ p_0(t_0) ~ d \log t_0 , \eqno(A5)
$$
as all moments of this distribution are finite.  

Now we shall convolve the results obtained above with 
a three-dimensional Gaussian distribution adopted for the lens velocities:
$$
p (V) ~ dV = \left( 2 \over \pi \right) ^{1/2} ~ 
\exp \left( - { V^2 \over 2 \sigma _V^2 } \right) ~ 
{ V^2 ~ \over \sigma _V ^2 } ~
{ dV \over \sigma _V } ,     \eqno(A6)
$$
where $\sigma_V$ is the one-dimensional velocity dispersion,
related to the three-dimensional velocity dispersion by
$\sigma _V=V_{rms}/\sqrt{3}$. The total number of events expected from
the lenses with such a distribution of velocities is given by
$$
N_\delta = \left( {9 \pi \over 32}\right)^{1/2} ~ n_s \tau ~
{ \Delta t \over t_{ \sigma} } ,     \eqno(A7a)
$$
where 
$$
t_{ \sigma}=t_m(V=\sigma _V )= { 1 \over \sigma _V } ~
\left( { GMD_s \over c^2 } \right) ^{1/2} .  \eqno(A7b)
$$
and the subscript $\delta$ indicates that 
all the lenses have identical masses, i.e., the mass function is a 
$\delta$ function.

After some algebra, the probability distribution of event time scales 
can be written as
$$
p_\delta(t) = {1 \over 2} \int_0^\infty \exp(-{v^2 \over 2})~v^4 ~ p_T(vt) ~dv, 
\hskip 1.0cm t = { t_0 \over t_{ \sigma} } ,  \hskip 1.0cm
v = { V \over \sigma _V } ,   \eqno(A8)
$$
where $p_T$ was defined with eq. (A3a).
The asymptotic behaviors are given by
$$
P_\delta(<t) = { 2^{11/2} \over 3 \pi^{3/2} } ~ t^3, \hskip 1.0cm
{\mathrm{for}} ~~ t \ll 1 ,  \eqno(A9a)
$$
$$
P_\delta(>t) = { 2^{11/2} \over 45 \pi^{3/2} } ~ {1 \over t^3}, \hskip 1.0cm
{\mathrm{for}} ~~ t \gg 1 .  \eqno(A9b)
$$

Now let us further adopt a general power law distribution of the lensing
masses:
$$
p (M) ~ dM \propto M^{ \alpha } ~ dM ,
\hskip 1.0cm {\mathrm{for}} \hskip 0.5cm M_{\mathrm{min}} \leq M \leq M_{\mathrm{max}}.
\eqno(A10)
$$
The normalized mass function can be rewritten in a logarithmic form:
$$
p ( \log m ) ~ d \log m = \left( { \ln 10 \over C( \alpha ) } \right)
~  m^{\alpha + 1} ~ d \log m , \hskip 0.5cm m \equiv { M \over M_0 } ,
\hskip 0.5cm M_0 \equiv ( M_{\mathrm{min}} M_{\mathrm {max}} )^{1/2} ,
\eqno(A11)
$$
where
$$
C( \alpha )=\int_{m_{\mathrm {min}}}^{m_{\mathrm{max}}} m^{ \alpha } dm =
\left\{		\begin{array}{ll}
		\ln(m_{\mathrm{max}}/m_{\mathrm{min}}), & \mbox{if $ \alpha =-1$} \\
		{(m_{\mathrm{max}}^{1+ \alpha }-m_{\mathrm{min}}^{1+ \alpha }) /
		(1+ \alpha )}, & \mbox{otherwise}
		\end{array}
\right.
\eqno(A12)
$$

After convolving with the mass function, the total number of
events is given by:
$$
N = \left({9 \pi \over 32}\right)^{1/2} n_s \tau ~ 
{\Delta t \over t_{\sigma,0}}
{ C(\alpha +0.5) \over C(\alpha +1)} , 
\eqno(A13)
$$
where $\tau$ is the total optical depth by all the lenses with different
masses and velocities, and the duration is normalized to
$$
t_{\sigma,0}=t_\sigma (M=M_0)=
{ 1 \over \sigma _V } ~
\left( { GM_0 D_s \over c^2 } \right) ^{1/2} .  \eqno(A14)
$$
The event duration distribution is given by
$$
p(t) = { 1 \over C(\alpha +0.5) }
~ \int_{m_{\mathrm{min}}}^{m_{\mathrm{max}}} m^{\alpha}~p_\delta(m^{-1/2} t)~dm ,
\hskip 1.0cm t \equiv { t_0 \over t_{ \sigma , 0 } } .
\eqno(A15)
$$
The asymptotic behaviors are given by
$$
P(<t) = { 2^{11/2} \over 3 \pi^{3/2} }
{C(\alpha -1) \over C(\alpha +0.5)} ~ t^3, \hskip 1.0cm
{\mathrm{for}}~ t\ll 1 ,  \eqno(A16a)
$$
$$
P(>t) = { 2^{11/2} \over 45 \pi^{3/2} }
{C(\alpha +2) \over C(\alpha +0.5)} ~ {1 \over t^3},
\hskip 1.0cm {\mathrm{for}}~ t \gg 1 .  \eqno(A16b)
$$
It is convenient to use a logarithmic probability distribution.
A few examples of such distribution functions are shown in Figs. 1 and 3.

De R\'ujula et al (1991)
proposed to use the moment analysis to deduce the distribution of
lens masses. After some algebra the first two moments can be
expressed as
$$
<\log t> = <\log t>_\delta+ {1 \over 2} \phi_1   ,    \eqno(A17a)
$$
$$
< ( \log t )^2> = < (\log t )^2 >_\delta + \phi_1 <\log t>_\delta + {1 \over 4} \phi_2 ,
\eqno(A17b)
$$
$$
\sigma^2_{\log t}=\sigma^2_\delta + {1 \over 4} (\phi_2 - \phi^2_1),
\hskip 1.0cm \sigma_\delta = 0.240 ,
\eqno(A17c)
$$
where the quantities with a subscript $\delta$ were
defined with eq. (5), and
$$
\phi_1 = 
{ m_{\mathrm{max}}^{1.5+\alpha} ( \log m_{\mathrm{max}} )
- m_{\mathrm{min}}^{1.5+\alpha} ( \log m_{\mathrm{min}} )
\over 
m_{\mathrm{max}}^{1.5+\alpha} - m_{\mathrm{min}}^{1.5+\alpha}} 
- {1 \over (1.5+\alpha) \ln 10} ,
\hskip 1.3cm \alpha \ne -1.5 , \eqno(A18a)
$$
$$
\phi_2 = 
{m_{\mathrm{max}}^{1.5+\alpha} ( \log m_{\mathrm{max}} )^2 
- m_{\mathrm{min}}^{1.5+\alpha} ( \log m_{\mathrm{min}} )^2
\over 
m_{\mathrm{max}}^{1.5+\alpha} - m_{\mathrm{min}}^{1.5+\alpha}} 
- {1 \over (1.5+\alpha) \ln 10} ~ \phi_1,
\hskip 0.5cm \alpha \ne -1.5 , \eqno(A18b)
$$
and
$$
\phi_1 = 
0.5 ~ \log ~ (m_{\mathrm{min}} m_{\mathrm{max}}),
\hskip 6.0cm \alpha = -1.5 , \eqno(A19a)
$$
$$
\phi_2 = 
{ 1 \over 3 } ~
[ ( \log m_{\mathrm{min}} )^2 + ( \log m_{\mathrm{min}} )( \log m_{\mathrm{max}} ) 
+ ( \log ,_{\mathrm{max}})^2 ],
\hskip 0.5cm \alpha = -1.5. \eqno(A19b)
$$

{}

\newpage

\begin{figure}
\epsscale{0.8}
\plotone{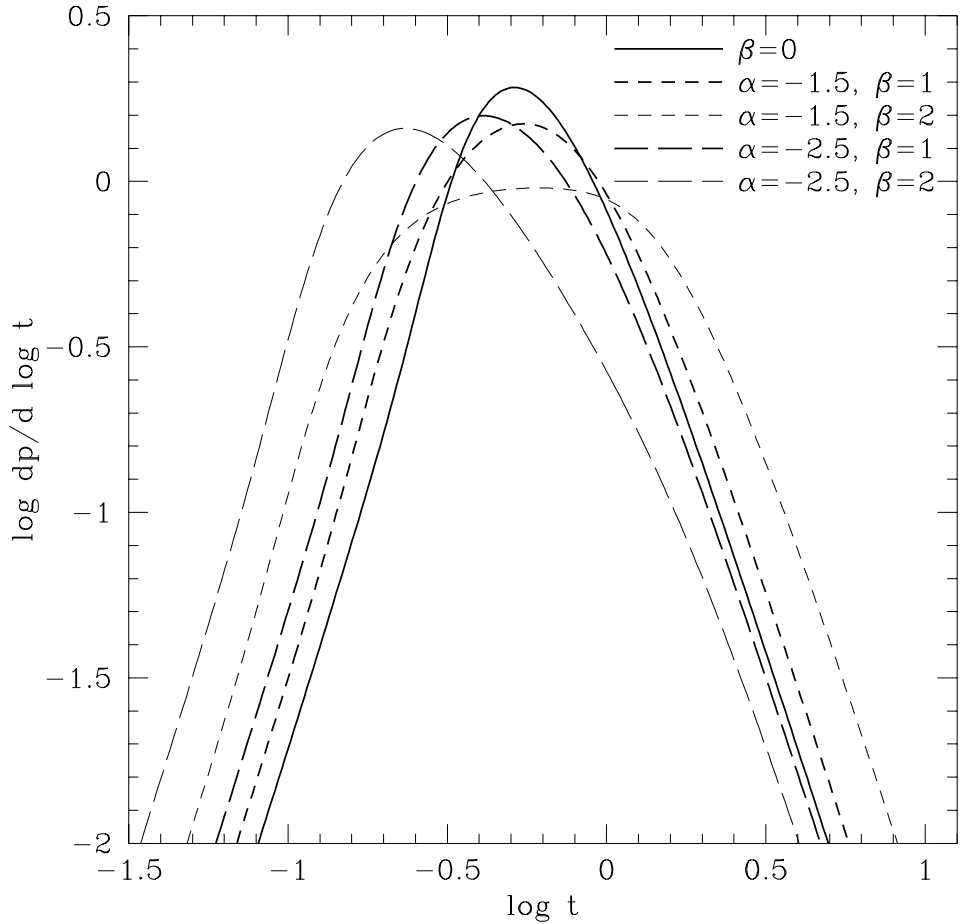}
\caption{
The probability distribution functions for the logarithm of event time
scale are shown for four power law mass functions:
$ ( \alpha , \beta ) = (-1.5,1), ~ (-1.5,2), ~ (-2.5,1), ~ (-2.5,2) $
where $ \alpha $ is the slope of the
mass function and $ \beta \equiv \log ~ (M_{\mathrm{max}} / M_{\mathrm{min}}) $.
All the time scales are in units of $t_\sigma $
corresponding to $ M = M_0 \equiv
(M_{\mathrm{min}} M_{\mathrm{max}})^{1/2} $ (cf. eqs. 2 and 6).
Also shown is the case for $ \beta = 0 $, which corresponds to
all lenses having the same mass (cf. eq. 4).
}
\end{figure}

\newpage
\begin{figure}
\epsscale{0.8}
\plotone{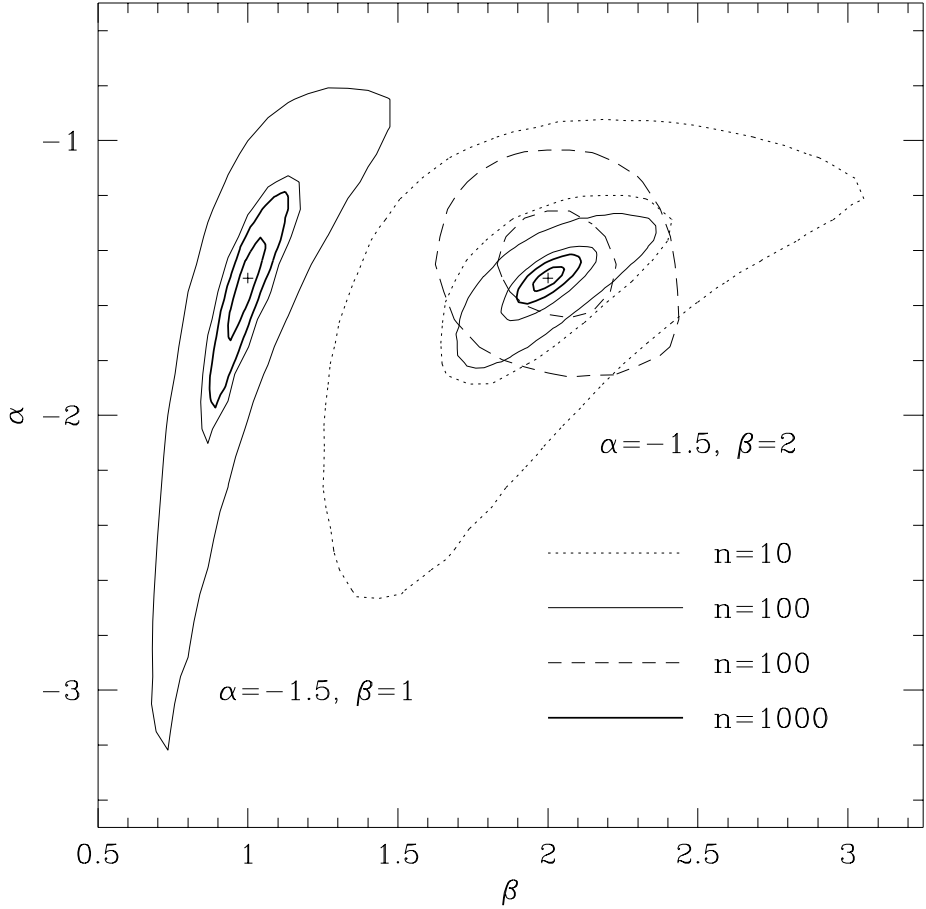}
\caption{
The contours corresponding to one and two standard deviations
are shown for different models of the mass function in the
$ \alpha - \beta $ plane, where $ \alpha $ is the slope of the
mass function and $ \beta \equiv \log ~ (M_{\mathrm{max}} / M_{\mathrm{min}}) $.
The models correspond to
$ ( \alpha , \beta ) = (-1.5,1), ~ (-1.5,2)$ as indicated by the crosses.
The contours are for an average number of microlensing events
$ n = 10 $ (dotted lines),
$ n = 100 $ (thin solid lines), and $ n = 1,000 $ (thick solid lines). 
The dotted and solid lines are for models with the mass scale set by fixing
the upper mass limit, i.e., declaring it to be known.
The dashed contours are for $( \alpha , \beta ) = (-1.5,1)$ with $n=100$
and with the mass scale set by the first moment (average) of the event
durations.   In all these simulations we assume that the observers 
can detect every microlensing event
with the impact parameter smaller than its Einstein ring radius,
i.e., that the detection system is perfect.
}
\end{figure}

\newpage
\begin{figure}
\epsscale{0.8}
\plotone{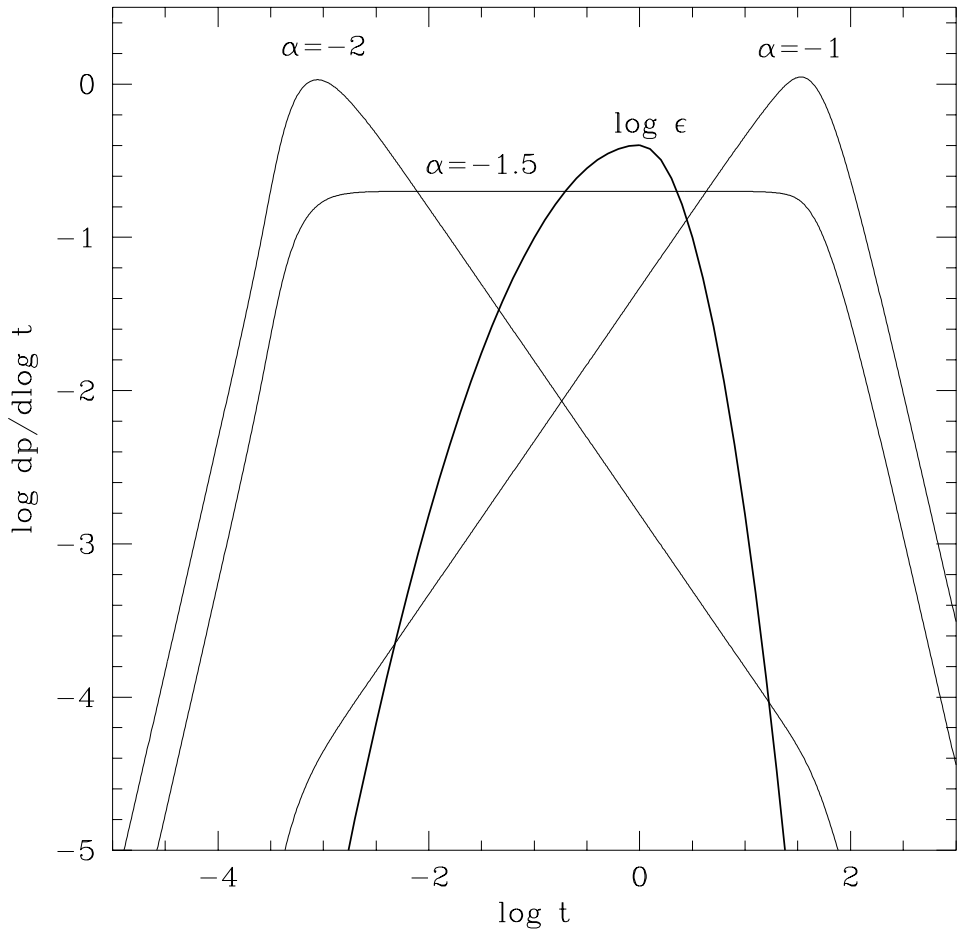}
\caption{
The logarithmic probability distributions for event time scales
are shown for three very broad mass functions with $ \beta = 10 $
($\beta_{\mathrm{min}}=-6, \beta_{\mathrm{max}}=4$),
and the slopes $ \alpha = -2, ~ -1.5, ~ -1 $, respectively.
The time scales are normalized to the duration at the peak
sensitivity (cf. eq. 10). Also 
shown is the approximate shape of the efficiency curve of the MACHO
and OGLE searches.  Notice, that in this case the majority of lensing
events are not detected as they fall outside of the relatively narrow
detectability window.
}
\end{figure}

\newpage
\begin{figure}
\epsscale{0.8}
\plotone{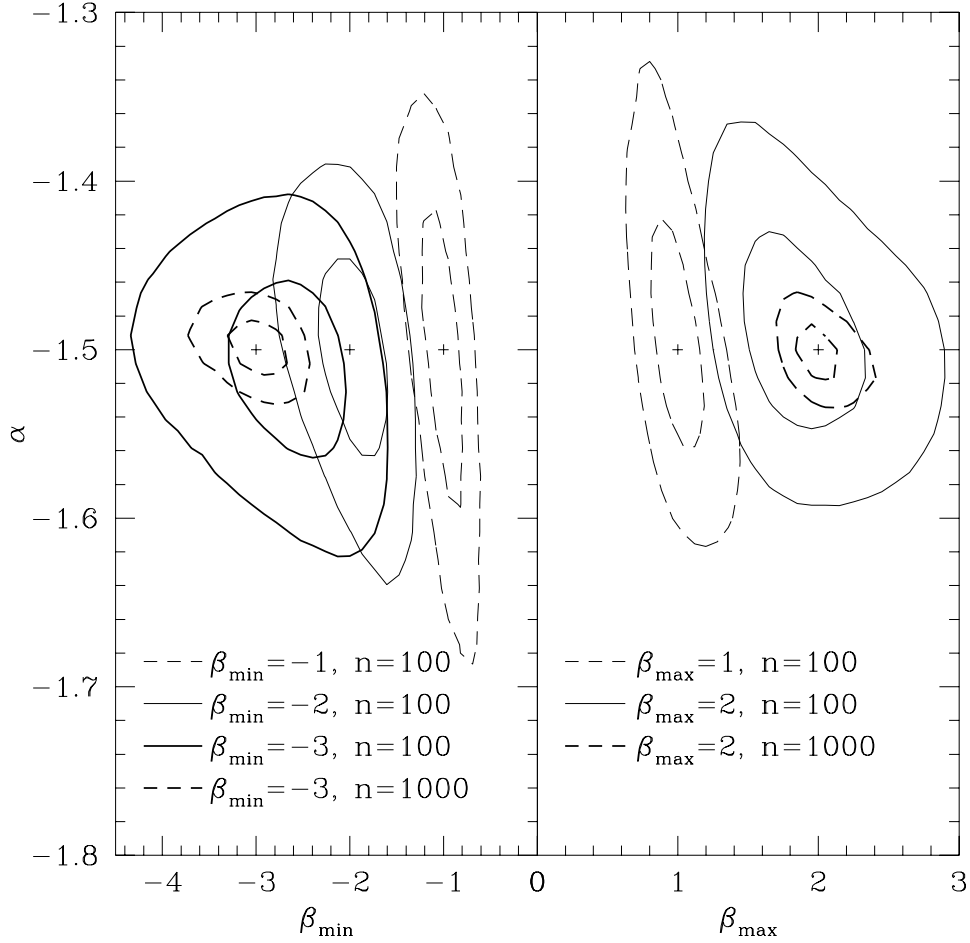}
\caption{
The contours corresponding to one and two standard deviations
and an average number of microlensing events $ n = 100 $ and $n=1000$
are shown in the $ \alpha - \beta $ plane, where $\alpha=-1.5$ is
the slope of the model lens mass function. For clarity, only two sets of
contours for $n=1000$ are shown. The models on the left correspond to
$ \beta _{\mathrm{max}} \equiv \log ~ (M_{\mathrm{max}} / M_p) \gg 1 $, i.e., the longest
events are beyond the detectability window as given with eq. (10)
and shown in Fig. 3.  The contours are for the mass functions with
$ \beta _{\mathrm{min}} \equiv \log ~ (M_{\mathrm{min}} / M_p ) = ~ -3, ~ -2, ~ -1 $.
The models on the right correspond to $ \beta _{\mathrm{min}} \ll -1 $, i.e.,
the shortest events are beyond the detectability window.  The 
contours are for the mass functions with $ \beta _{\mathrm{max}} = ~ 1, ~ 2 $.
}
\end{figure}

\end{document}